\documentclass{article}
\usepackage{spconf,amsmath,graphicx, amssymb, color, subcaption}
\usepackage[ruled,vlined]{algorithm2e}

\usepackage{cite, soul}

\title{DCT-based air interface design for function computation}

\name{Marc Martinez-Gost$^{1, 2}$ \qquad Ana Pérez-Neira$^{1, 2, 3}$ \qquad Miguel Ángel Lagunas$^{1}$
\thanks{This work is part of the project IRENE (PID2020-115323RB-C31), funded by MCIN/AEI/10.13039/501100011033.}
}

\address{$^{1}$ Centre Tecnològic de Telecomunicacions de Catalunya, Spain \\
$^{2}$ Dept. of Signal Theory and Communications, Universitat Politècnica de Catalunya, Spain\\
$^{3}$ ICREA Acadèmia, Spain\\
}

\begin{document}
%
\maketitle
\begin{abstract}
With the integration of communication and computing, it is expected that part of the computing is transferred to the transmitter side. In this paper we address the general problem of Frequency Modulation (FM) for function approximation through a communication channel. We exploit the benefits of the Discrete Cosine Transform (DCT) to approximate the function and design the waveform. In front of other approximation schemes, the DCT uses basis of controlled dynamic, which is a desirable property for a practical implementation.  Furthermore, the proposed modulation allows to recover both the measurement and the function in a single transmission. Our experiments show that this scheme outperforms the double side-band (DSB) modulation in terms of mean squared error (MSE). This can also be implemented with an agnostic receiver, in which the function is unknown to the receiver. Finally, the proposed modulation is compatible with some of the existing transmission technologies for sensor networks.
\end{abstract}


\section{Introduction}
\label{sec:intro}
Future networks are expected to integrate computing capabilities and will consist of many devices capable of sensing the environment and taking decisions. In this vein, \cite{computing_in_networks} highlights the relevance of computing in communication networks and motivates it with respect to a wide variety of use cases.

Both terrestrial an non-terrestrial wireless sensor networks (WSN) \cite{vetterli, Wu2021} will play a key role, together with Mobile Edge Computing (MEC) \cite{mec, Kim2022}. As latency can be a bottleneck in these networks for some kind of services (e.g., disaster relief communications, maritime surveillance, etc.), part of the computation will be displaced to the transmitter side in order to reduce the amount of transmitted data, latency, energy consumption, among many other metrics. Thus, the system has to find a trade off between the offloaded and the locally processed data \cite{satellite_iot, Merluzzi_2021, master, leyva2022}.


The emergence of joint communication and computing schemes can be traced back to Gastpar et al. \cite{vetterli}, where they show that when the goal of the communication is not to transmit information reliably, but to assist the function computation of the data, the source-channel separation theorem does not hold in general. In \cite{uncoded_optimal} the author shows that analog uncoded transmission is optimal to estimate Gaussian uncorrelated sources through a Gaussian multiple-access channel. In terms of mean squared error (MSE), the distortion achieved by this scheme scales exponentially better with the number of measurements than when using a digital scheme.
These initial information theoretic studies boosted the development of analog schemes for joint communication and computing, as for Over-the-Air Computing (AirComp) \cite{comp_mac, aircomp_review, nomographic_wsn}.

In this work we deal with the general problem of Frequency Modulation (FM) for function approximation through a communication channel,
in which the figure of merit is the MSE between the original function value and its estimation at the receiver.
This paper focuses on the simplest study case, which is considering a unique measurement at the transmitter and a unique function to be computed in a point-to-point setting. Nevertheless, it is shown that the proposed scheme is promising for distributed scenarios as in \cite{nomographic_wsn}. This study reveals that the use of the Discrete Cosine Transform (DCT) for function representation has advantages in the design of joint communication and computing schemes as, for instance, those needed in WSNs. Although the Internet of Things (IoT) and machine to machine (M2M) connectivity landscape is fragmented, the relevance of FM is demonstrated by its use, for instance, in Sigfox 
or in LoRa
\cite{LoRa_modulation, review_phy}. Our proposed scheme naturally integrates FM and it is the first one that studies an FM scheme for function approximation that allows to use current IoT or M2M transports. We emphasize that sensors demand a very simple technology and this scheme provides a light weight processing without compromising the precision of the computation.

We focus on scenarios that require computing the function at the transmitter instead of the receiver. Therefore, the reconstruction error always exhibits a linear relationship with the channel noise. Our scheme allows the receiver to recover both, the measurement and the function value in a single transmission, and there is the possibility that the receiver does not need the specific knowledge of the function to be computed; thus, being an agnostic receiver in this case.

It is crucial not to confuse this scheme with the use of the DCT for data compression. Although the DCT has been used in distributed scenarios to compress data and reduce the communication burden \cite{dct_wsn}, in this work we only deal with one measurement, and the DCT is used to approximate its function representation. To the best of our knowledge, this is the first work devoted to the study and design of modulation schemes for joint communication and computing schemes.

The remaining part of the paper proceeds as follows: Section 2 begins by laying out the theoretical dimensions of the research and presents the novel waveform for computing; Section 3 is concerned with the analytical performance of the scheme and the benchmarks; Then, Section 4 presents some experiments to complement the theoretical findings and Section 5 concludes the paper.

\section{Methodology}
\label{sec:method}

\subsection{Signal Model}
Consider a signal $x(t)$, which could be associated to a measurement from a sensing environment (e.g., temperature) or a value resulting from a local computation (e.g., distributed learning model). We assume an observation time of $T$ over which $x(t)$ remains constant. For instance, in a sensing environment $T$ may be associated to the changing nature of the parameter, whereas in a computation scenario to the time it takes to acquire and process data. Thus, throughout the rest of the paper we drop the time index and work with $x$.

We assume a point-to-point setting where the transmitter acquires $x$ and uses a quantizer $Q: x \rightarrow m$ to discretize $x$ into a finite set of $N$ levels as
\begin{equation}
Q(x) = \Delta\biggl\lfloor \frac{x}{\Delta} +\frac{1}{2}\biggr\rfloor=m
\end{equation}
where $\lfloor\cdot\rfloor$ is the floor function and $\Delta$ is the step size. Without loss of generality, we assume $x$ to be normalized in the range $[0, N-1]\subset\mathbb{R}$ and, correspondingly, $m\in [0, N-1]\subset\mathbb{N}$, which imposes $\Delta=1$. Throughout the rest of the paper we refer to $m$ as the measurement.

A function $f(m)$ of the measurement is to be available at the receiver. For that, the datum $m$ is processed at the transmitter with the DCT for function approximation, then modulated and sent over the channel. For the latter we consider an additive white Gaussian noise (AWGN) channel. At the receiver side, demodulation and processing is performed to obtain an estimate of the function $\hat{f}(m)$. The performance of the system is measured in terms of MSE between the original function and the reconstruction:
\begin{equation}
    MSE(f, K, m) = \big|f(m)-\hat{f}(m) \big|^2,
    \label{eq:loss}
\end{equation}
where $K$ parameterizes the DCT-based approximation quality of the function ${f}(m)$. In order to evaluate the performance over the whole regime of the measurement space, we define the average error as
\begin{equation}
    \overline{MSE}(f, K) =
    \frac{1}{N}\sum_{m=0}^{N-1}MSE(f, K, m)
    \label{eq:mean_loss}
\end{equation}

\begin{figure}[t]
\centering
\includegraphics[width=\columnwidth]{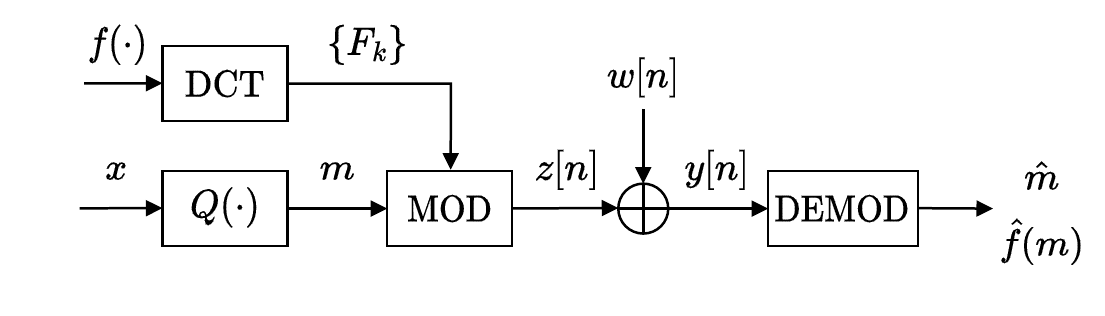}
\caption{Digital baseband scheme of the joint communication and computing scenario for the point-to-point system and function $f$.}
\label{fig:estimation_scheme}
\end{figure}

In (\ref{eq:loss}) we will take into account the effect of the DCT approximation and the channel noise, but not the quantization error. In this work we consider low-pass equivalent signals and a discrete-time model. In this respect, the bandwidth of signal $x$ is $W=1/T$ and we choose a sampling frequency of $f_s=N/T$, returning $N\geq2$ samples per measurement.
The channel is ideal with AWGN noise, $w$, with spectral density equal to $N_o/2\, [W/Hz]$. Fig. \ref{fig:estimation_scheme} shows the joint communication and computing scheme.

\subsection{DCT-based Function Approximation}
Consider a scalar univariate function $f$ to be approximated. We will adopt the Discrete Cosine Transform (DCT) because of the following advantages: First, its strong energy compaction property allows to reconstruct the function with only a small portion of the coefficients; second, the approximation error can be controlled by the number of disregarded coefficients due to its orthogonal basis; and third, its implementation complexity is $O(N\log(N))$, far below other function approximations, such as Volterra \cite{woods2002}.
Besides, in the following section we will see that the DCT brings further advantages in terms of communication.

For the set of function values $f(m)$ evaluated over the measurement space, we will use the Type-II DCT definition:
\begin{equation}
F_k = g_k\sum_{m=0}^{N-1} f(m) \cos\left(\frac{\pi k(2m+1)}{2N}\right),
\label{eq:dct}
\end{equation}
for $k=0,\dots, N-1$, where $g_0=1/\sqrt N$ and $g_k=\sqrt{2/N}$ otherwise. The $F_k\in\mathbb{R}$ are termed the DCT coefficients. Regarding the inverse DCT (iDCT), we use the definition provided by the orthonormal Type-III DCT:
\begin{equation}
f(m) = \sum_{k=0}^{N-1} g_kF_k \cos\left(\frac{\pi k(2m+1)}{2N}\right),
\label{eq:idct}
\end{equation}
for $m=0,\dots, N-1$. We define $\tilde{f}_K(m)$ to be the function approximated by preserving a percentage $\alpha$ of the energy, corresponding to the $K$ largest DCT coefficients $F_k$. We order $F_k$ in the set $\mathcal{K}_{\alpha}$ and $K=|\mathcal{K}_{\alpha}|$ is the size of the set. Throughout the rest of the paper we drop the subscript $\alpha$ and assume it is constant. Since not all the coefficients are retained, the function will not be perfectly reconstructed. In this respect we define (\ref{eq: error_dct}) as the approximation error in $f(m)$ due to the $K$-th DCT approximation.
\begin{align}
    MSE_{DCT}(f, & K, m) = \big|f(m)-\tilde{f}_K(m) \big|^2\nonumber\\
    &=\bigg|\sqrt{\frac{2}{N}}\sum_{k\not\in \mathcal{K}} F_k\cos\left(\frac{\pi k(2m+1)}{2N}\right)\bigg|^2
    \label{eq: error_dct}
\end{align}

Analogous to (\ref{eq:mean_loss}) we define (\ref{eq:mean_aprox_loss}) as the average $K$-th approximation error of $f$ over all measurements. Due to the orthogonality of the DCT basis, we can establish the reciprocal relationship between the preserved energy and the corresponding MSE. Notice that, with no consideration of the noise, this results in a lower bound on the minimum achievable error.
\begin{align}
    \overline{MSE}_{DCT}(f, K) = \frac{1}{N}\sum_{k\not\in \mathcal{K}} F_k^2=\frac{(1-\alpha)}{N}\sum_{k=0}^{N-1}F_k^2
    \label{eq:mean_aprox_loss}
\end{align}

\begin{figure}[t]
\centering
     \begin{subfigure}[b]{0.45\columnwidth}
         \includegraphics[width=\columnwidth]{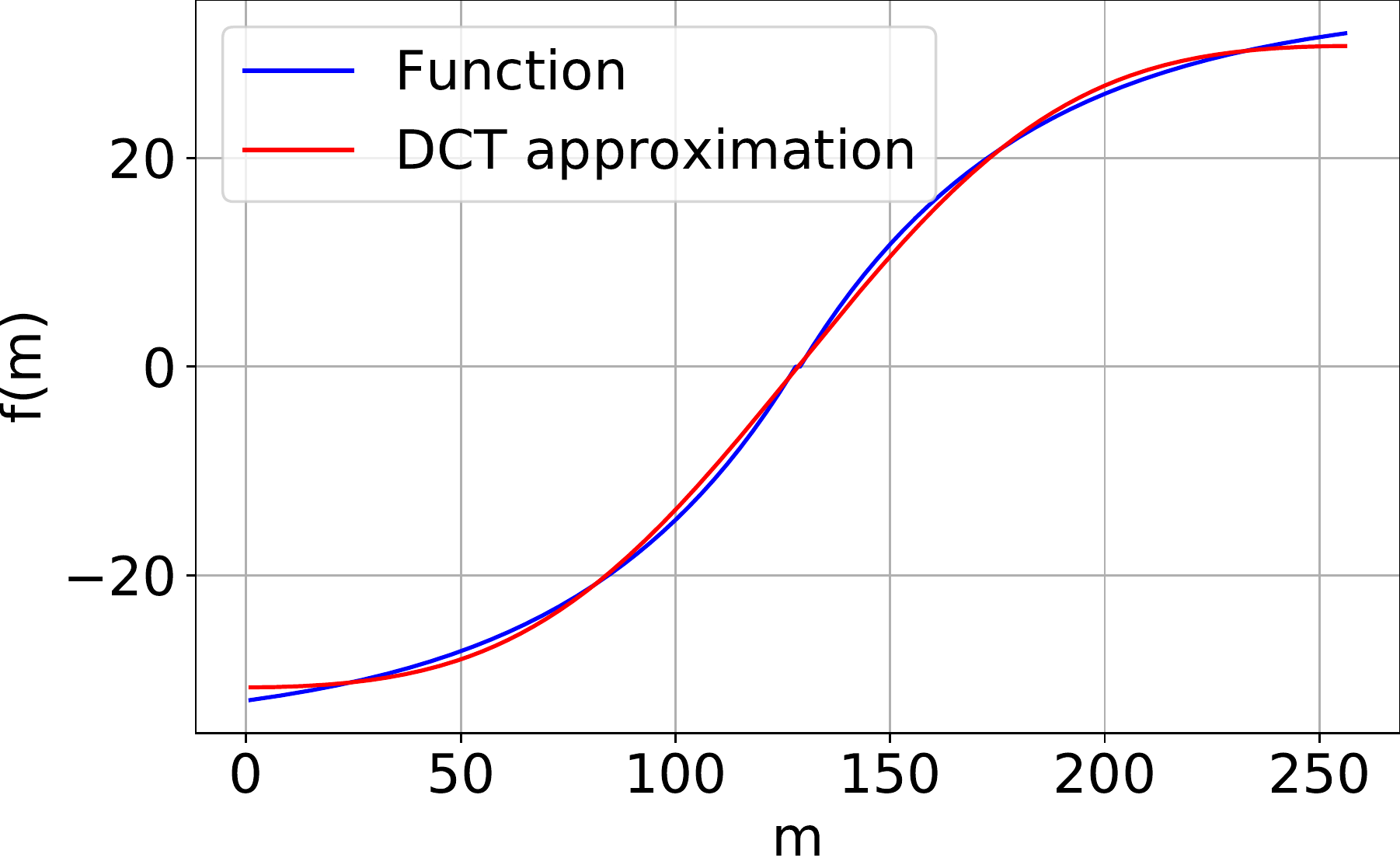}
         \caption[]{Sigmoid ($K=2$).} 
         \label{fig:sigmoid}
     \end{subfigure}
     \hspace{10pt}
     \begin{subfigure}[b]{0.45\columnwidth}
         \includegraphics[width=\columnwidth]{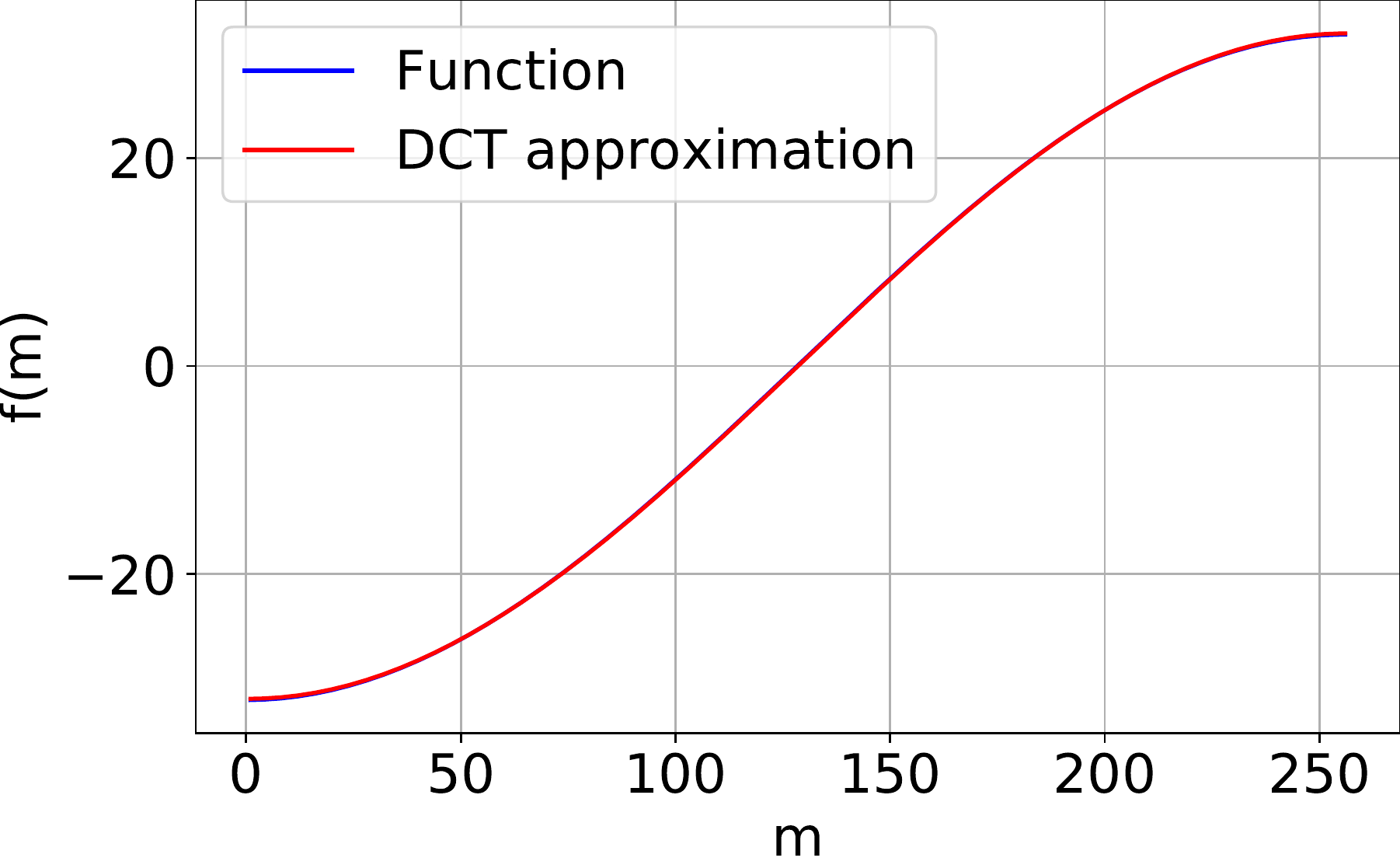}
         \caption[]{Sine ($K=1$).} 
         \label{fig:sine}
     \end{subfigure}

     \vskip\baselineskip
     \begin{subfigure}[b]{0.45\columnwidth}
         \includegraphics[width=\textwidth]{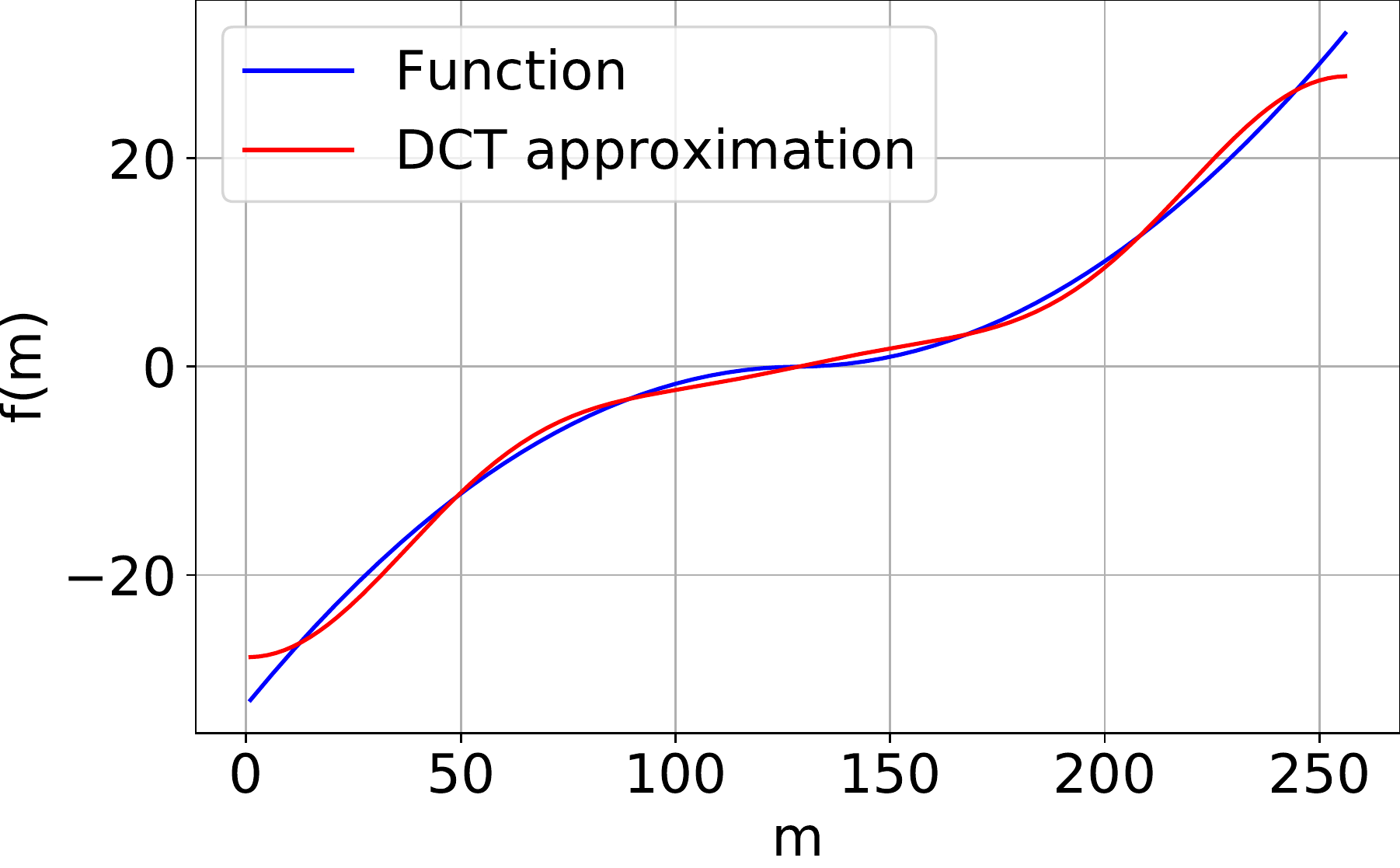}
         \caption[]{Square ($K=3$).} 
         \label{fig:square}
     \end{subfigure}
     \hspace{10pt}
     \begin{subfigure}[b]{0.45\columnwidth}
         \centering
         \includegraphics[width=\columnwidth]{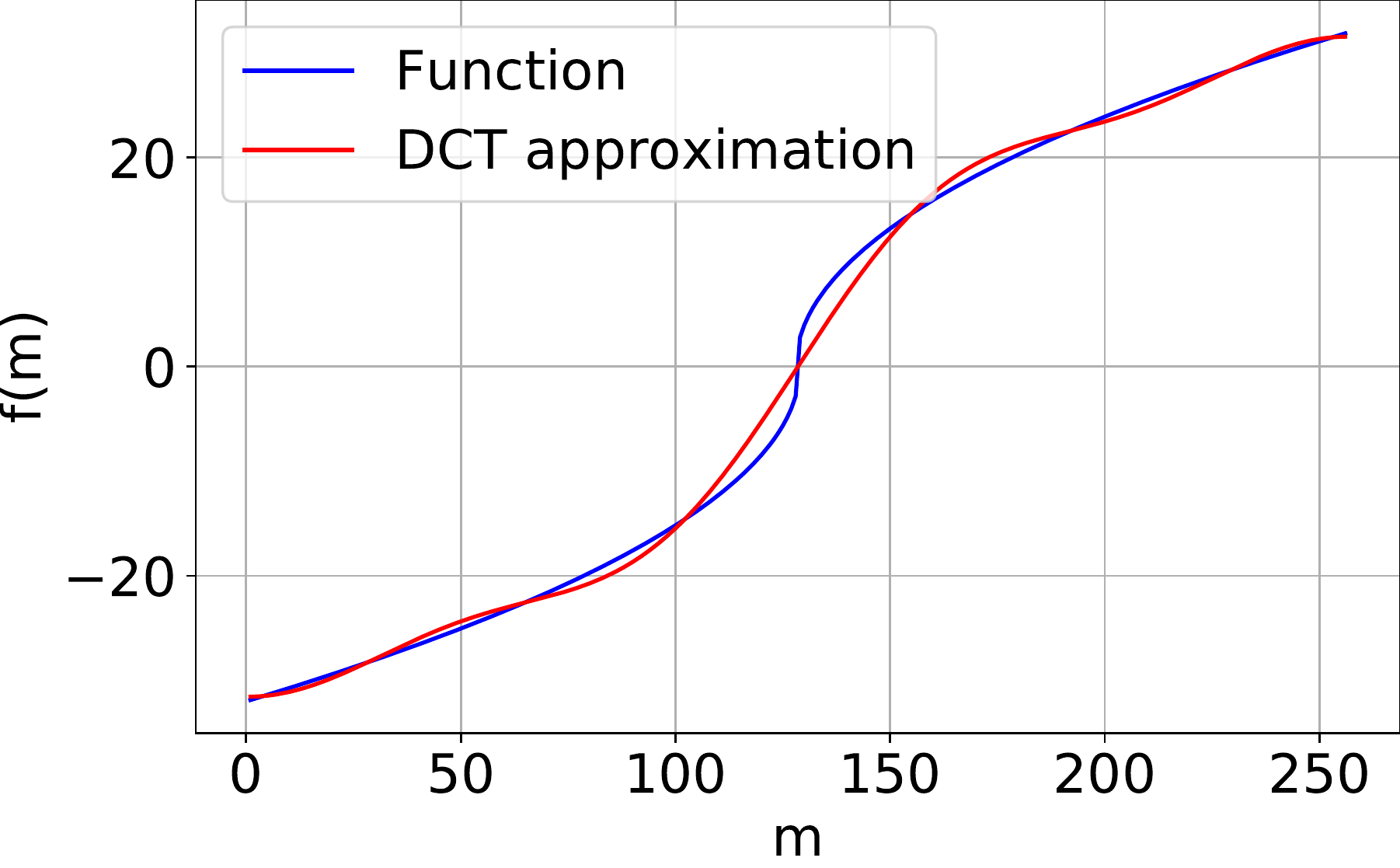}
         \caption[]{Square root ($K=3$).} 
         \label{fig:root}
     \end{subfigure}
        \caption{DCT approximation with $N=256$ for different functions  preserving $\alpha=0.995$ of the energy.} 
        \label{fig:dct_examples}
\end{figure}

Consider the function space $\mathcal{F}$ of smooth functions $f: \mathbb{R} \rightarrow \mathbb{R}$ with odd symmetry. Otherwise, any function in a given range can be redefined to have odd symmetry. For instance, Fig. \ref{fig:square} shows $f(m)=f_{max}sign(m-N/2)(m-N/2)^2$ with $f_{max}=1/2N$, which is the odd version of the square function. This geometry allows higher compression because only the odd coefficients are retained, independently of $f$. This results in the following definitions:
\begin{align}
    K=& \min\bigg\{k\in\mathbb{N}: \sum_{i=0}^{2k-1}F_k^2\geq \alpha\sum_{i=0}^{N-1}F_k^2\bigg\}\label{eq:K}\\
    \mathcal{K}=& \{2k-1: k\leq K\}\label{eq:set_K}
\end{align}

Besides, the elements in $\mathcal{K}$ are ordered in decreasing magnitude. Fig. \ref{fig:dct_examples} shows the DCT approximation for different functions to preserve $\alpha=99.5\%$ of the energy, which only requires between 2 and 3 coefficients per function.

\subsection{DCT-FM: Modulation for Computing}
We identify that the iDCT can be used as a waveform if we incorporate a time index. Consider the measurement $m$ and the set of DCT coefficients $F_k$ for $k\in\mathcal{K}$ of function $f$. The samples of the proposed modulation at baseband, which we term DCT-FM, are
\begin{equation}
z[n] = A_c\sqrt{\frac{2}{N}}\sum_{k\in \mathcal{K}} F_k \cos\left(\frac{\pi k(2m+1)}{2N}n\right),
\label{eq:dct_mod}
\end{equation}
for $n=0,\dots, N-1$, where $A_c$ is the amplitude of the carrier. This modulation is generated by sampling the measurement signal at $f_s$, returning $N$ samples of the constant measurement $m$. Notice that (\ref{eq:dct_mod}) is built with $K$ FM tones whose frequency depends on the index of the DCT coefficient $k$ and the measurement $m$. Since the frequencies of the tones in (\ref{eq:dct_mod}) are orthogonal, the iDCT applied over the time signal $z[n]$, results in the coefficients $F_k$ located at frequencies governed by $k$ and $m$.
We highlight that this is a digital transmission of an analog message (i.e., the quantizer can have an arbitrary large number of levels) and $N$ samples are transmitted for the same measurement $m$.

Notice that other function approximations may be accurate enough in terms of computing, but not suitable for the purpose of communication. For instance, the Volterra approximation \cite{volterra} uses powers of the input signal, which does not limit the dynamic range of the signal (i.e., the transmission power). Conversely, the nature of the DCT-FM limits the amplitude of the signal and preserves communication bandwidth. The bandwidth of the baseband transmitted signal in (\ref{eq:dct_mod}) is shown in (\ref{eq: bw_dct}) for measurement $m$, whereas the maximum bandwidth, $B_z^{max}$, is associated to the maximum measurement, i.e., $m=N-1$. The price to pay for joint communication and computing is an increasing bandwidth and the transmission of $K$ simultaneous carriers. The more coefficients, the smaller the MSE, but the more bandwidth expansion. Thus, there is a trade off between the approximation accuracy and the bandwidth. This is true, however, if the transmitted coefficients are above the noise level.
\begin{equation}
    B_z=\frac{(2K-1)(2m+1)}{4}W
    \label{eq: bw_dct}
\end{equation}


\subsection{DCT-FM demodulation}
After downconversion, filtering at $B_z^{max}$ and sampling at $f_s$, the signal at the input of the receiver is
\begin{equation}
y[n] = z[n] + w[n]\quad \text{for} \quad n=0,\dots, N-1,
\label{eq:signal_rx}
\end{equation}
where $w$ corresponds to complex AWGN samples with power $2N_oB_z^{max}$. We consider perfect carrier and phase synchronization at downconversion so that $y[n]\in\mathbb{R}$.

\subsubsection{Measurement estimate}
\label{sec:measurement_estimate}
The receiver computes a iDCT over the time samples $n$. Since the transmitted signal is modulated in frequency, it will obtain $K'$ peaks, as seen in (\ref{eq:demod}). The noise level determines the number of detectable tones, which is why $K'\neq K$ and $\mathcal{K}'\subseteq \mathcal{K}$.
\begin{align}
 r[l]&=\frac{1}{A_c}\text{iDCT}\{y\}[l]\nonumber\\
 &=
 \begin{cases} 
      \tilde{F}_k=F_k + \frac{1}{A_c}\tilde{w}[l] & l=\text{ifreq}(k,m) \\
      \frac{1}{A_c}\tilde{w}[l] & \text{otherwise}
   \end{cases}
\label{eq:demod}
\end{align}
where the noise $\tilde{w}$ remains Gaussian with power $\sigma^2=N_o/2\, [W]$, index $l$ is the frequency component of the iDCT and ifreq$(k,m)$ corresponds to the frequency bin associated to each transmitted tone. The relationship between these peaks and $m$ can be found analytically matching the frequency of each transmitted tone in (\ref{eq:dct_mod}) and the discrete frequency after applying the iDCT, which corresponds to 
\begin{equation}
\frac{1}{2}\frac{1}{2N}k(2m+1)=
\frac{\text{ifreq}(k,m)-1/2}{2N}\quad \forall k\in\mathcal{K}',
\end{equation}
where the additional $1/2$ term in the right-hand sided numerator comes from the iDCT frequencies starting at 1 and not $1/2$. Moreover, the frequency is normalized dividing it by the maximum frequency index, which is $2N$. This can be rewritten as
\begin{equation}
\hat{m} = \frac{\text{ifreq}(k,m)-(k-1)/2}{k}
\quad \forall k\in\mathcal{K}'
\label{eq:match_freqs}
\end{equation}

It is reasonable to estimate $\hat m$ from the fundamental harmonic, i.e., $k=1$, because it is associated to the largest DCT coefficient and has the largest signal-to-noise ratio (SNR). From there, expression (\ref{eq:match_freqs}) can be used sequentially to find the indices associated to the harmonics, $\text{ifreq}(k,m)$, and extract their respective amplitudes $\tilde{F}_k$. For each one of the harmonics, its power is compared to a threshold $\sigma_{th}^2$, defined in terms of the noise level. Since the coefficients are scanned in decreasing power, once a frequency component is below the threshold, the algorithm stops. The gain provided by this scheme is equivalent to setting to zero all the components of (\ref{eq:demod}) which do not contain signal information. Thus, only the noise components in $\mathcal{K}'$ are retained. The design of an efficient threshold $\sigma_{th}^2$ that maximizes the probability of detection is out of the scope of this paper. One choice is to provide a false alarm probability $P_{FA}$ on the first coefficient. This corresponds to $\sigma_{th}^2=\sigma^2 Q^{-1}(P_{FA})$, where $Q^{-1}(\cdot)$ is the inverse $Q$-function and the details are omitted for brevity. Furthermore, observe that this algorithm is function-agnostic as it only needs to know the function set $\mathcal{F}$ to know $\mathcal{K}$, but not which specific function is being computed.

It is noteworthy to mention that for $\text{ifreq}(k,m)>N$, the harmonics move backwards in a cyclic way, meaning that the $\text{ifreq}(k,m)$ does not necessarily corresponds to the transmitted frequency. The overall algorithm to locate the transmitted frequency components is described in Algorithm \ref{alg: peak_search}.

Fig. \ref{fig:peaks_sigmoid_M2_m40_SNR_15} shows the computation of the iDCT at the receiver for the sigmoid function using $K=3$ coefficients, $m=60$ and $SNR=15\ dB$. Using (\ref{eq:match_freqs}), the measurement is estimated correctly at $\hat{m}=60$. With the same expression, $\text{ifreq}(2,40)=180$ and $\text{ifreq}(3,40)=300$. While the former satisfies $l=\text{ifreq}(2,40)$, the third tone exceeds $N$. Using the Algorithm \ref{alg: peak_search}, this tone is located at index $l=212$. However, this tone is below the noise level and not detectable, meaning that the reconstruction error is lower bounded by $MSE_{DCT}(f,2,m)$. This is, at most, and considering no effect of the noise, the function will be approximated using only 2 coefficients, not 3.


\begin{algorithm}[t]
\DontPrintSemicolon
\SetAlgoLined
\KwIn{ $r$, $\sigma^2_{th}$ }
\vspace*{4 pt}
\KwOut{ $\hat{m}, \mathcal{K}', \{F_k\}_{k\in\mathcal{K}'}$ }
\vspace*{4 pt}
$k=1$\;
$\mathcal{K}'=\{k\}$\;
$\hat{m} = \text{argmax } |r|$\;
$\tilde{F}_k = r[\hat{m}]$\;
keep $\gets$ True\;
\vspace*{2pt}
 \While{\upshape keep}{
    $k\gets k+2$\;
    $ifreq = k\ \hat{m} + (k-1)/2$\;
    \uIf{$\bigl\lfloor ifreq/N \bigr\rfloor \text{\upshape mod } 2==0$}{
        $l= ifreq \text{ mod } N$ \;
      }
      \Else{
        $l= N - (ifreq \text{ mod } N)$ \;
      }
    \uIf{$|r[l]|^2>\sigma^2_{th}$}{
          $\mathcal{K}'=\mathcal{K}'\cup k$\;
          $\tilde{F}_k = r[l]$\;
      }
      \Else{
        keep $\gets$ False \;
      }
 }
 \caption{Peak Search}
 \label{alg: peak_search}
\end{algorithm}

\subsubsection{Function estimate}
The function can be reconstructed using $\tilde{F}_k$ and $\hat m$ via the iDCT. Notice that this transformation differs from the previous one: the iDCT in (\ref{eq:demod}) happens in the time domain, i.e., index $n$, whereas this iDCT is computed in the measurement space, i.e., index $m$.
The function value can be estimated as
\begin{align}
\hat{f}(m) &= \sqrt{\frac{2}{N}}\sum_{k\in\mathcal{K}'} \tilde{F}[k] \cos\left(\frac{\pi k(2m+1)}{2N}\right)
\label{eq:estimate}
\end{align}

\begin{figure}[t]
\centering
\begin{subfigure}[b]{\columnwidth}
   \includegraphics[width=\columnwidth]{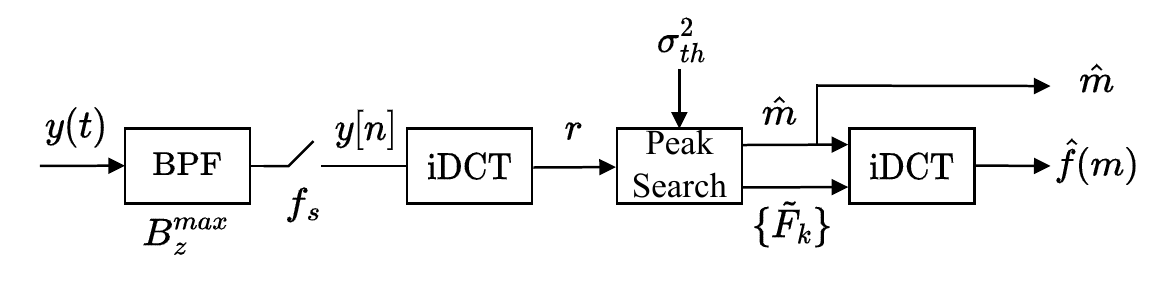}
   \caption{}
   \label{fig:receiver_scheme} 
\end{subfigure}

\begin{subfigure}[b]{\columnwidth}
   \includegraphics[width=0.9\columnwidth]{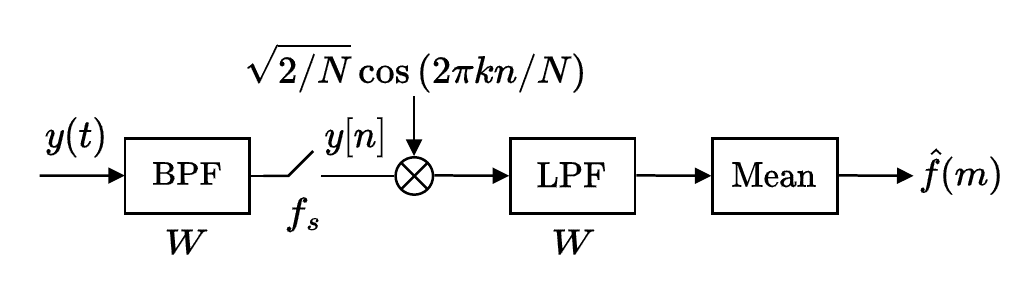}
   \caption{}
   \label{fig:receiver_scheme_DSB}
\end{subfigure}

\caption{Receiver scheme to estimate $\hat{f}(m)$: (a) DCT-FM, and (b) DSB.}
\end{figure}

Fig. \ref{fig:receiver_scheme} shows the overall signal processing workflow deployed at the receiver to recover both, the measurement and the function estimate.

On the other hand, see that $z[1]=f(m)$, this is, the function can be estimated from a single sample. This provides an even simpler demodulation scheme, although it requires high SNR.

\subsection{Modulation properties}
The presented modulation is an FM scheme designed to communicate a computed function on a measurement.
In this paper we study the basic point-to-point scheme, however it is also a promising system for scenarios with distributed transmitters that may take advantage of AirComp.
While in this work $z[n]$ is generated by a unique transmitter, it could have been generated by $K$ transmitters in a distributed fashion (i.e., assuming perfect synchronization at reception). We leave the study of these distributed schemes for future work.


Frequency-based modulations and access schemes are prevalent in current M2M communication systems (e.g., OFDM, FSK, etc.), together with phase modulations. The energy compaction of the DCT is adequate for multiple access schemes since it reduces the interference a user can generate. In this respect, we highlight the LoRa modulation \cite{LoRa_modulation}, which is a FSK modulation using a chirp spread spectrum (CSS). The DCT-FM presented in this paper allows the integration with the LoRa modulation, as the transmitted signal in (\ref{eq:dct_mod}) can be extended by means of a chirp. By using the analytic signal of $z[n]$,
\begin{equation}
z_{an}[n] = A_c\sqrt{\frac{2}{N}}\sum_{k\in \mathcal{K}} F_k \exp\left(\frac{j\pi k(2m+1)}{2N}n\right),
\label{eq:dct_mod_analytic}
\end{equation}
the resulting LoRa modulation is
\begin{equation}
    z_{LoRa}[n] = z_{an}[n]\exp{\left(j\frac{\pi f_{mod}}{N}n^2\right)},
    \label{eq:lora_mod}
\end{equation}
for $n=0,\dots,N-1$, where $f_{mod}$ allows to control the spreading factor of the chirped signal. While this spreading over the spectrum does not offer any gain in terms of noise, it is adequate for multiple access schemes that need to be robust against interferences. Besides, the same LoRa receiver can be used where the FFT would substitute the first iDCT that is indicated in Fig. \ref{fig:receiver_scheme}.

Regarding the use of analog FM \cite{caarlson}, there are two issues that hinder its integration into communication systems for computing, which motivates the present work. First, with respect to our scheme, analog FM can modulate either the measurement or the function, but not both so that the receiver can obtain them simultaneously.
Second, the receiver needs to implement a FM descriminator (e.g., phase estimator), which is not compatible with the existent IoT transmission systems.
In conclusion, we propose a digital frequency modulation scheme for analog data that overcomes these aspects and can trade-off bandwidth and quality at reception as analog FM does.

\begin{figure}[t]
\centering
\includegraphics[width=\columnwidth]{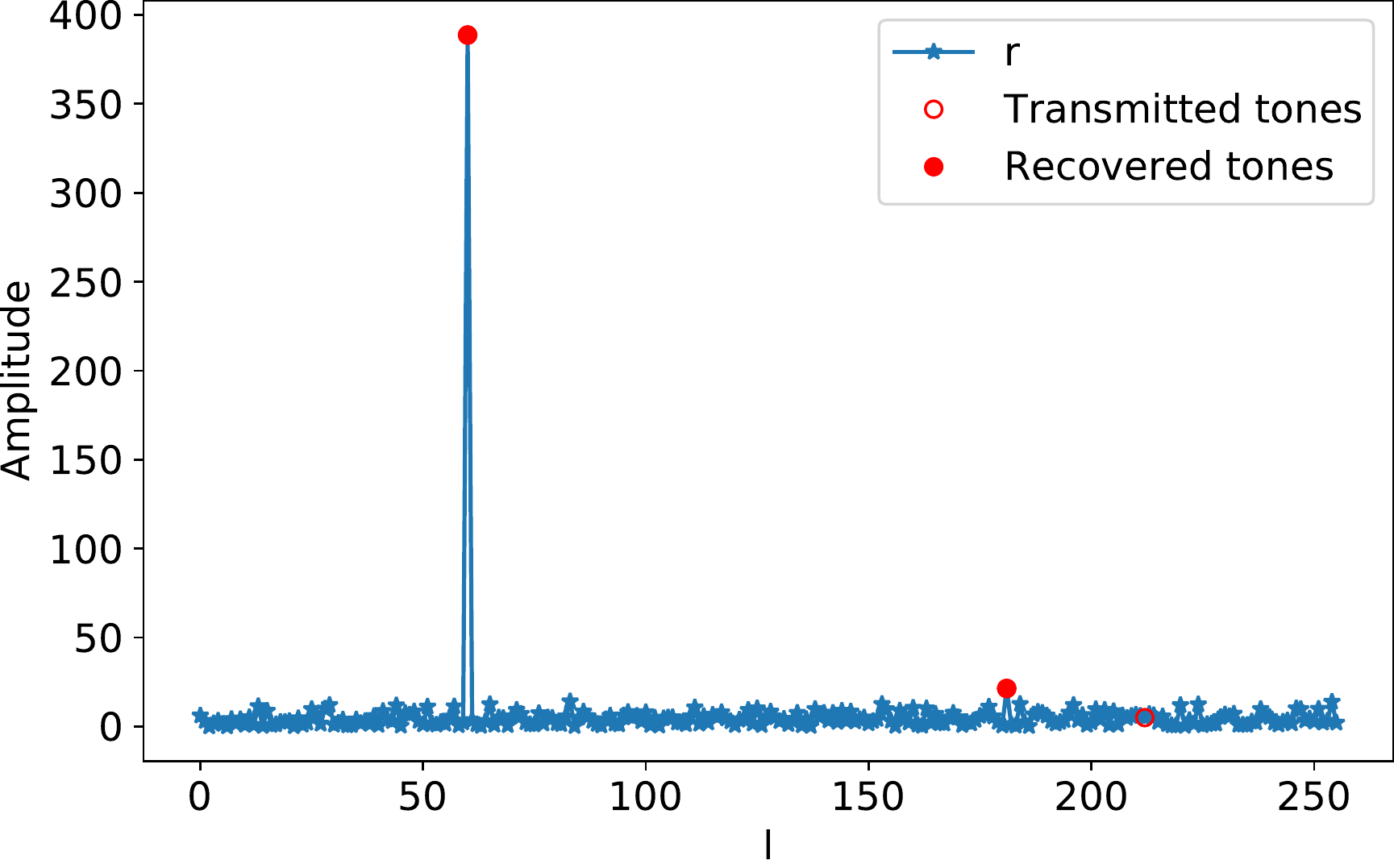}
\caption{Signal $r$, iDCT at the receiver for the sigmoid function with $K=3$, $m=60$ and $SNR=15\ dB$.} 
\label{fig:peaks_sigmoid_M2_m40_SNR_15}
\end{figure}

\section{Analysis of the reconstruction error}


We distinguish between two cases: in the former we will study the performance of the proposed modulation in section \ref{sec:method}, referred as the agnostic receiver. In the latter we perform a slight variation in the modulation for which the receiver is no longer agnostic, although it enhances the performance of the scheme.

\subsection{Agnostic receiver}
The power of the transmitted signal in (\ref{eq:dct_mod}) is
\begin{equation}
P_{ag} = \frac{1}{N}\sum_{n=0}^{N-1}|z[n]|^2=\frac{A_c^2}{N}\sum_{k\in\mathcal{K}}|F_k|^2,
\label{eq: snr}
\end{equation}

We assume that $P_{ag}$ is large enough with respect to the noise for $m$ to be always correctly detected from the fundamental frequency. Thus, in the rest of the paper we assume $\hat m=m$. By substituting (\ref{eq:demod}) in (\ref{eq:estimate}), the function estimate can be expressed as
\begin{equation}
\hat{f}(m) = {\tilde f}_{K'}(m) + v(m),
\label{eq:f_estimate}
\end{equation}
where ${\tilde f}_{K'}(m)$ is the $K'$-th DCT approximation of $f(m)$ and the noise term $v(m)$ is
\begin{align}
v(m) &= \frac{1}{A_c}\text{iDCT}\{\tilde{w}\}[m]\nonumber\\
&=\sqrt{\frac{2}{NA_c^2}}\sum_{k\in\mathcal{K}'}\tilde{w}[k] \cos\left(\frac{\pi k(2m+1)}{2N}\right)
\end{align}

Notice that we still assume $g_k=\sqrt{2/N}$. The detection of $k=0$ as a transmitted frequency can be rejected systematically because no function is generated with a DC component. Since $\mathcal{K}'$ as well as $m$ are known at the receiver by Algorithm \ref{alg: peak_search}, the overall distribution is Gaussian as $v(m)\sim\mathcal{N}\left(0, \rho(m)\sigma^2\right)$ with
\begin{equation}
    \rho(m) = \frac{2}{NA_c^2}\sum_{k\in\mathcal{K}'}\cos^2\left(\frac{\pi k(2m+1)}{2N}\right)
\end{equation}

The MSE of the agnostic receiver scheme (\ref{eq:f_estimate}) is shown in (\ref{eq:mse_dct}), where the first term corresponds to the error due to DCT approximation.
\begin{align}
    MSE_{ag}(f, K, m) &= MSE_{DCT}(f,K',m) +\rho(m)\sigma^2
    \label{eq:mse_dct}
\end{align}
and the average MSE results in
\begin{align}
    \overline{MSE}_{ag}(f, K) = \overline{MSE}_{DCT}(f,K')+\frac{K'\sigma^2}{NA_c^2}
    \label{eq:mse_mean}
\end{align}

Notice that \eqref{eq:mse_mean} explicitly shows a joint dependence on the communication (i.e., transmitted power) and computation (function approximation).
One of the benefits of moving the function computation at the transmitter is that the noise affects linearly to the reconstruction error. The term $K'$ in the numerator evinces the number of noise samples that are not filtered out in the function reconstruction. Also, as expected, the higher the number of samples $N$, the lower the effect of the channel noise.

\subsection{Non-agnostic receiver}
\label{sec:nag}
In the modulation presented in (\ref{eq:dct_mod}) the power of the DCT coefficients decay very fast. As seen in Fig. \ref{fig:peaks_sigmoid_M2_m40_SNR_15}, the power of the tones decay around $12$ dB/coefficient. In this respect, we perceive a threshold effect which prevents the receiver from properly detecting beyond two or three coefficients, unless working with a higher SNR. In here we present a slight variation of the previous modulation, which allows to detect more harmonics and we term it as the non-agnostic receiver.

Consider the waveform
\begin{equation}
z[n] = A_c^{nag}\sqrt{\frac{2}{N}}\sum_{k\in \mathcal{K}} \frac{1}{2^{\frac{k-1}{2}}} \cos\left(\frac{\pi k(2m+1)}{2N}n\right),
\label{eq:dct_mod_non_agnostic}
\end{equation}
for $n=0,\dots, N-1$, which corresponds to the same frequency modulation as in (\ref{eq:dct_mod}) without considering the DCT coefficients. Conversely, each tone is transmitted with half of the power from the previous tone.
The transmitted power of the non-agnostic DCT-FM is
\begin{equation}
P_{nag} =\frac{(A_c^{nag})^2}{N}\sum_{k\in\mathcal{K}}2^{1-k},
\label{eq: snr_nag}
\end{equation}
from which we can define the amplitude of the carrier to maintain the same transmitted power level as in the agnostic scheme:
\begin{equation}
    P_{ag}=P_{nag} \Rightarrow
    A_c^{nag} = A_c\sqrt{\frac{\sum_{k\in\mathcal{K}}F_k^2}{\sum_{k\in\mathcal{K}}2^{1-k}}}
    \label{eq:amplitude_nag}
\end{equation}

Thus, the advantage of this power distribution is twofold: First, this allows to maintain more tones above the noise level because it is distributed more uniformly; and second, the tones are still ordered in decreasing amplitude, meaning that the demodulation process described in section \ref{sec:measurement_estimate} can be still used to recover an estimate of the measurement and recover the number of transmitted tones. By substituting the amplitudes of the harmonics with the corresponding DCT coefficients, the receiver can obtain an estimate of the function value.

In terms of error, the advantage of this scheme is that the coefficients are noiseless because they are imposed by the receiver. Thus, the unique source of error is the DCT approximation over the $K'$ detected tones. Nevertheless, this is achieved at the expenses of not having an agnostic receiver, as it has to know the function of interest to insert the DCT coefficients. Note that a unique tone would be enough for the receiver to recover $m$ and approximate the function with the iDCT at the desired accuracy, since the receiver has access to the coefficients. However, for fairness to the agnostic scheme, we consider the case in which the receiver uses as many coefficients as detected tones.

The error produced by this non-agnostic receiver scheme is
\begin{align}
    MSE_{nag}(f,K,m) = MSE_{DCT}(f,K',m),
    \label{eq:mse_non_agnostic}
\end{align}
and the average MSE results in
\begin{align}
    \overline{MSE}_{nag}(f, K) = \overline{MSE}_{DCT}(f,K')
    \label{eq:mse_mean_non_agnostic}
\end{align}

Notice that, for a given SNR level, the number of detected tones $K'$ is not necessarily the same as in the agnostic receiver. The non-agnostic receiver becomes a detection, not an estimation problem, since the number of detected peaks $K'$ determines the accuracy of the reconstruction.

\subsection{Benchmarking}
We note that in the current literature \cite{aircomp_review}, when these problems are studied, they generally assume that baseband signals add up coherently in the channel, which is only achievable via linear analog modulations.
Specifically we study the double sideband (DSB) transmission, because its coherent demodulation provides a better performance than a standard amplitude modulation (AM).

Consider the following sampled baseband DSB modulation,
\begin{equation}
z_{DSB}[n] = A_c^{DSB}\sqrt{\frac{2}{N}}\tilde{f}_K(m)\cos{\left(\frac{2\pi kn}{N}\right)},
\label{eq:AM-mod}
\end{equation}
for $n=0,\dots, N-1$, where $k/N$ is the discrete frequency. The transmitted power of this scheme is
\begin{equation}
P_{DSB} = \frac{(A_c^{DSB})^2}{N}|\tilde{f}_K(m)|^2
\label{eq: snr_dsb}
\end{equation}

To maintain the same transmission power as the DCT-FM, we can define $A_c^{DSB}$ as
\begin{equation}
    P_{ag}=P_{DSB} \Rightarrow
    A_c^{DSB}(m) = \sqrt{\frac{A_c^2\sum_{k\in\mathcal{K}}F_k^2}{|\tilde{f}_K(m)|^2+\epsilon}},
    \label{eq:amplitude_dsb}
\end{equation}
where $\epsilon$ prevents the denominator from vanishing when $\tilde{f}_K(m)\approx 0$. Note that, in this way, $A_c^{DSB}$ depends on the measurement $m$. The bandwidth of (\ref{eq:AM-mod}) is $1/T=W$ and, thus, the noise at the receiver has the same power as $\tilde{w}$ in (\ref{eq:demod}). After bandpass filtering and sampling at $f_s$, the receiver uses a coherent demodulator, returning $N$ noisy samples of the amplitude. The minimum mean square error (MMSE) estimator, which coincides with the Maximum Likelihood (ML) estimator (i.e., the average mean), is used to reduce the effect of the noise. The overall receiver is depicted in Fig. \ref{fig:receiver_scheme_DSB}. The MSE of DSB is shown in (\ref{eq:mse_am}), whose development is omitted for brevity.
\begin{align}
    MSE_{DSB}(f,K,m) = &\ MSE_{DCT}(f,K,m) \nonumber\\&+\,  \frac{\tilde{f}_K(m)^2\sigma^2}{A_c^2\sum_{k\in\mathcal{K}}F_k^2}\quad
    \label{eq:mse_am}
\end{align}

Unlike (\ref{eq:mse_dct}), the DCT approximation error is computed over $\mathcal{K}$, not $\mathcal{K}'$, since the whole computation is performed at the transmitter.



Expression (\ref{eq:mse_am_mean}) shows the average error of DSB, which presents the same behavior as the agnostic scheme: at high SNR both schemes are dominated by the DCT approximation; while at low SNR, the DSB modulation outperforms the proposed scheme by a factor of $2K'$, as seen in (\ref{eq:gap_mean}). Using the Cauchy-Schwarz inequality, it follows that $MSE_{ag}(f,K,m)\geq MSE_{DSB}(f,K,m)$, whose details are omitted for brevity. Comparing both transmission power expressions, namely (\ref{eq: snr}) and (\ref{eq: snr_dsb}), the power assigned to the tones in DCT-FM is independent of the measurement, whereas in DSB the allocation depends on $m$. Thus, for a small function value $\tilde{f}_K(m)$ that would be below the noise level by itself, DSB assigns more power to the carrier via $A_c^{DSB}$. This makes it robust against the noise, which explains why the $MSE_{DSB}$ depends linearly with the power of the function value.

Nevertheless, the non-agnostic DCT-FM of section \ref{sec:nag} outperforms both schemes because it produces a pure frequency modulated signal. 
Conversely, the agnostic DCT-FM and DSB carry information in the amplitude of the signal, in which the noise has a larger impact. 
\begin{align}
    \overline{MSE}_{DSB}(f, K)
    = \overline{MSE}_{DCT}(f,K)+\frac{\sigma^2}{2NA_c^2}
    \label{eq:mse_am_mean}\\
    K'=K \Rightarrow
    \overline{MSE}_{DSB}(f, K)=\frac{1}{2K'}\overline{MSE}_{ag}(f, K)
    \label{eq:gap_mean}
\end{align}

\section{Results}
\subsection{Functions of interest}
These functions highly depend on the case of study, although some of them are of general interest. For instance, the use of companding is very frequent in communication systems, which can be implemented using a sigmoid function (see Fig. \ref{fig:sigmoid}), although a very cheap version can be implemented using a sine function (see Fig. \ref{fig:sine}). This function is very suitable for the DCT-FM scheme, as it only requires 1 coefficient for a perfect reconstruction.
In the case of learning models and WSN data fusion, linear, square, truncation and saturation functions are standard building blocks (e.g., batch normalization).

\subsection{Experimental setup}
In this section we will provide experimental evidence of the performance offered by the previous modulations, namely:
\begin{itemize}
    \item \textbf{DCT-ag}: This corresponds to the agnostic receiver scheme, with average performance as (\ref{eq:mse_mean}) plus the effect of the peak search algorithm.
    \item \textbf{DCT-nag}: This corresponds to the non-agnostic receiver scheme, with average performance as (\ref{eq:mse_mean_non_agnostic}).
    \item \textbf{DSB}: This corresponds to the linear DSB modulation, with average performance as (\ref{eq:mse_am_mean}).
    \item \textbf{Mean MSE-ag}: This corresponds to the agnostic receiver scheme, with side information $K$ and always recovering $K$ coefficients. Its average performance is like (\ref{eq:mse_mean}).
\end{itemize}

The effect of the peak search algorithm in DCT-ag cannot be neglected. Its threshold level determines the number of coefficients $K'$ that will be used to estimate the function, which translates in introducing more noise in the estimate. Notice in (\ref{eq:mse_mean}) that a new coefficient reduces the DCT approximation error, but increases the noise component. In order to visualize the effect of this scheme without the effect of the peak search, we also analyze the Mean MSE-ag, which has no influence of the thresholding. This algorithm assumes that the receiver has knowledge on the transmitted tones $K$, and it exactly uses $K'=K$ coefficients to estimate the function.

In the following we will test their average performance for different transmission powers. We set the minimum transmission power to $-5\ dB$ because it is the minimum at which the fundamental can be recovered in the agnostic modulation for the considered dynamic range ($|f(m)|\leq32$). Nevertheless, the transmission power may not represent a limitation as these networks (e.g., WSN) are generally hierarchical. This allow to control the number of hops and helps to overcome the limitations that the used modulation may have in terms of SNR. Finally, the noise power is constant and set to $0\ dB$.

We study the sigmoid function, although the results are similar for the other ones. We assume an observation time of $T=1\ s$, $N=256$ samples and $\alpha=0.995$ (i.e., $K=3$). Regarding the DCT-FM we assume $A_c=1$, and we set a threshold of $\sigma_{th}^2=8\sigma^2$, which corresponds to a probability of false alarm of $10^{-16}$. The experiments are averaged over $10^2$ Monte Carlo runs and the MSE is normalized by the energy of the function value.

\subsection{Performance analysis}

\begin{figure}[t]
\centering
\includegraphics[width=\columnwidth]{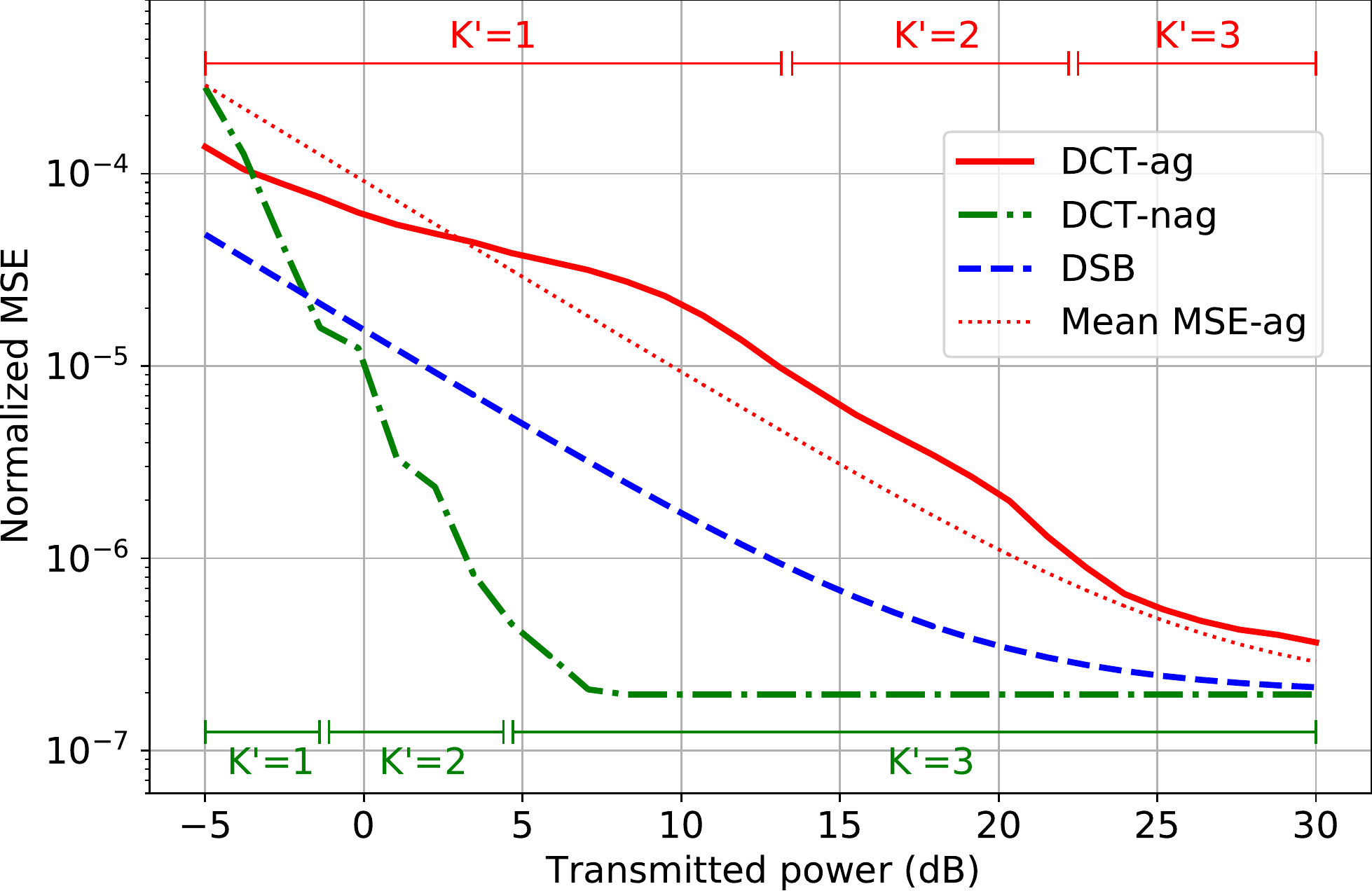}
\caption{Average error, $\overline{MSE}(f,K)$, for the sigmoid function with $K=3$ coefficient.}
\label{fig:sim_sigmoid_M3_mean}
\end{figure}

\begin{figure}[!t]
\centering
\includegraphics[width=\columnwidth]{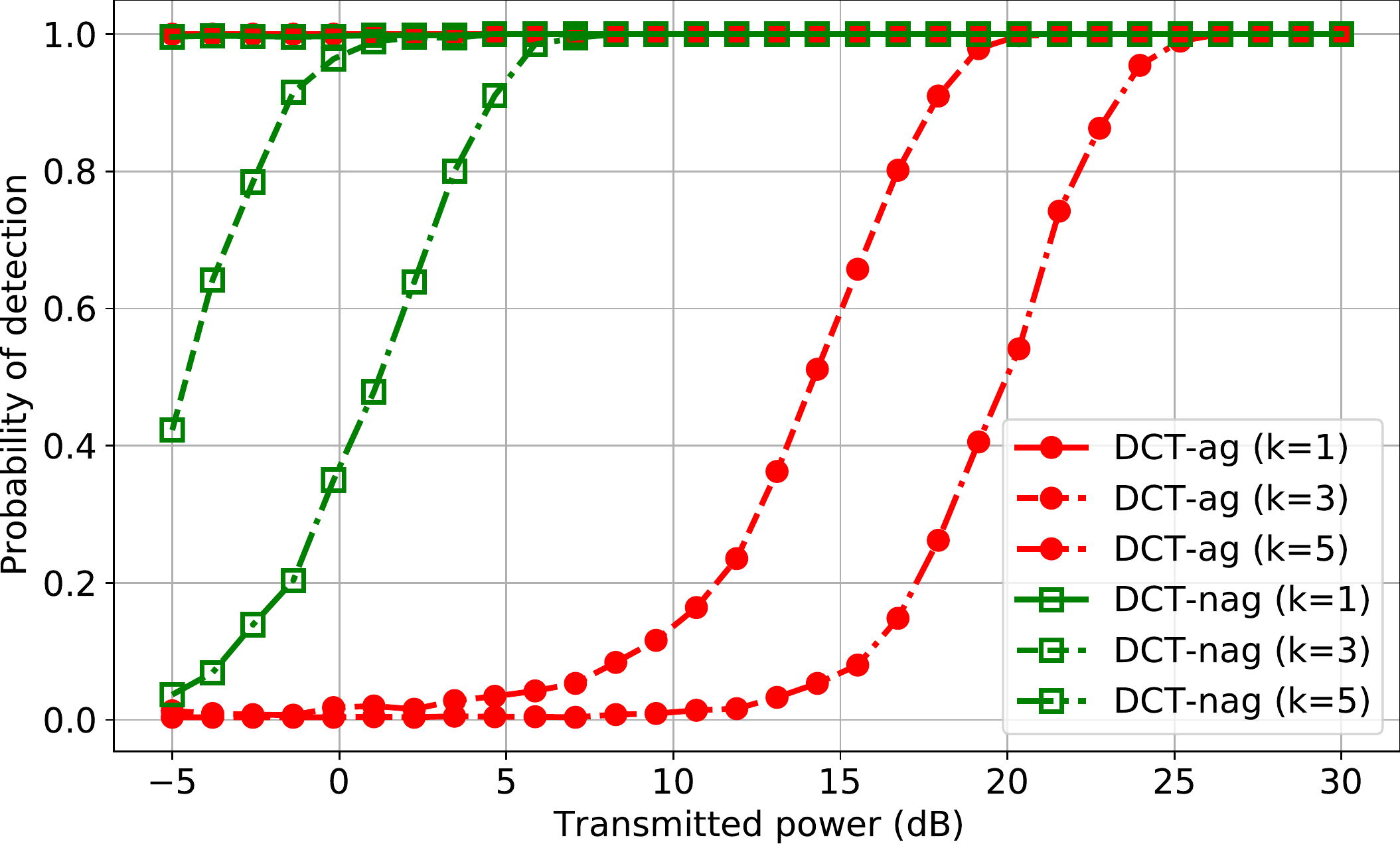}
\caption{Probability of detection per coefficient for the scenario in Fig. \ref{fig:sim_sigmoid_M3_mean}.}
\label{fig:sim_sigmoid_M3_mean_coeffs}
\end{figure}

Fig. \ref{fig:sim_sigmoid_M3_mean} shows $\overline{MSE}(f,K)$ versus the transmission power.
The DCT-nag outperforms the other modulations up to $-3\ dB$. Likewise, Fig. \ref{fig:sim_sigmoid_M3_mean} highlights the power regimes at which the DCT-ag and DCT-nag detect $K'$ coefficients with a $90\%$ probability. This is, the DCT-ag starts missing the third coefficient at $22.5\ dB$, whereas the DCT-nag can recover it up to $5\ dB$. This is associated to the power distribution among the tones. In this respect, Fig. \ref{fig:sim_sigmoid_M3_mean_coeffs} shows the probability of detection for each of the three coefficients and for both DCT-FM schemes. For the same coefficient there is gap of 15 dB in performance. Below $-5\ dB$, the DCT-ag starts missing the fundamental coefficient, whereas the DCT-nag can still recover up to 2 coefficients. This also explains why there is a gap between DCT-ag and Mean MSE-ag: At $-5\ dB$ DCT-ag can only recover one coefficient, while the Mean MSE-ag has side information $K=3$ and always uses 3 coefficients. Nevertheless, for $k=\{3,5\}$, the samples are mostly noise and they worsen the reconstruction, as seen in Fig. \ref{fig:sim_sigmoid_M3_mean}.


\section{Conclusions}
Analog modulations are the optimal transport for joint communications and computing purposes, where AM or DSB are the de-facto modulations that are used in the literature, i.e., for Over the Air Computing or distributed learning models. This paper presents the FM counterpart to the DSB scheme, that relies on the DCT to approximate a function and transport the information simultaneously. The proposed scheme is compatible with LoRa and allows to recover the measurement and the function in a single transmission. It also supports the implementation of an agnostic receiver to the computed function, while the non-agnostic receiver outperforms the former and the benchmark in terms of MSE.
In the future we will expand the functionality of this modulation to specific computing applications via the nomographic representation of functions.

\bibliographystyle{IEEEtran}
\bibliography{refs}

\end{document}